\documentclass[%
 reprint,
 superscriptaddress,
 nofootinbib,
 nobibnotes,
 showpacs,
 amsmath,amssymb,
 aps,
 prl,
 longbibliography,
 floatfix,
]{revtex4-2}

\ExplSyntaxOn
\msg_redirect_name:nnn { siunitx } { physics-pkg } { none }
\ExplSyntaxOff
\usepackage{silence}
\usepackage{graphicx}
\usepackage{dcolumn}
\usepackage{bm}
\usepackage[mathlines]{lineno}

\usepackage{xspace}
\usepackage[version=4]{mhchem}
\usepackage{xfp}
\usepackage{circledsteps}
\usepackage{ulem}

\newcommand{\rangedash}{\raisebox{0.15ex}{\textendash}}
\newcommand{\hyphen}{\raisebox{0.15ex}{\text{-}}}
\newcommand{\SIrangeDash}[3]{%
  \num{#1}\rangedash\num{#2}~\si{#3}%
}
\newcommand{\SIhyphen}[2]{\num{#1}\hyphen\si{#2}}
\pgfkeys{
  /csteps/inner xsep=2.5pt,
  /csteps/inner ysep=2.5pt,
  /csteps/outer color=black,
  /csteps/inner color=black,
}

\usepackage{mathtools}

\usepackage[
  number-unit-product=~,
  per-mode=symbol,
  separate-uncertainty=true,
  range-units=single,
  range-phrase={--},
  list-units=single,
  list-final-separator={\text{, and }}
]{siunitx}
\usepackage{physics}
\AtBeginDocument{\RenewCommandCopy\qty\SI} %

\usepackage[utf8]{inputenc}
\usepackage[T1]{fontenc}
\usepackage{times}
\usepackage{etoolbox}
\newcommand{\subm}[1]{_{\mathrm {#1}}}

\newcommand{\thetaK}{\theta\subm{K}}

\newcommand{\thetaT}{\theta\subm{total}}

\newcommand{\DTK}{\Delta\theta\subm{K}}

\newcommand{\Neel}{N\'eel\xspace}
\newcommand{\TN}{T\subm{N}}

\newcommand{\FMO}{\ce{Fe2Mo3O8}\xspace}

\newcommand{\maintitle}{Unconventional Magneto-Optical Kerr Effect beyond Magnetization Scaling\\Driven by Pulse-Field-Induced Metamagnetism in \texorpdfstring{\FMO}{Fe2Mo3O8}}

\definecolor{myurlcolor}{HTML}{000080}
\usepackage[unicode,colorlinks=true,
  linkcolor=red, citecolor=red, urlcolor=myurlcolor
]{hyperref}

\begin{document}


\title{
	\maintitle
}

\author{Soichiro~Yamane}
\email{yamane.soichiro.a11@kyoto-u.jp}
\affiliation{Department of Electrical, Electronic, and Digital Science and Engineering, Kyoto University, Kyoto, Kyoto 615-8510, Japan}

\author{Sota~Nakamura}
\affiliation{Department of Electrical, Electronic, and Digital Science and Engineering, Kyoto University, Kyoto, Kyoto 615-8510, Japan}

\author{Shuya~Suzuki}
\affiliation{Department of Electrical, Electronic, and Digital Science and Engineering, Kyoto University, Kyoto, Kyoto 615-8510, Japan}

\author{Atsutoshi~Ikeda}
\affiliation{Department of Electrical, Electronic, and Digital Science and Engineering, Kyoto University, Kyoto, Kyoto 615-8510, Japan}

\author{Masaki~Kondo}
\affiliation{The Institute for Solid State Physics, The University of Tokyo, Kashiwa, Chiba 277-8581, Japan}

\author{Takashi~Kurumaji}
\affiliation{Division of Physics, Mathematics and Astronomy, California Institute of Technology, Pasadena, California 91125, USA}
\affiliation{Institute of Quantum Information and Matter, California Institute of Technology, Pasadena, California 91125, USA}

\author{Kentaro~Ino}
\affiliation{Department of Physics, Institute of Science Tokyo, Meguro-ku, Tokyo 152-8551, Japan}

\author{Keisuke~Matsuura}
\affiliation{Department of Physics, Institute of Science Tokyo, Meguro-ku, Tokyo 152-8551, Japan}

\author{Fumitaka~Kagawa}
\affiliation{Department of Physics, Institute of Science Tokyo, Meguro-ku, Tokyo 152-8551, Japan}
\affiliation{RIKEN Center for Emergent Matter Science, Wako, Saitama 351-0198, Japan}

\author{Yoshinori~Tokura}
\affiliation{RIKEN Center for Emergent Matter Science, Wako, Saitama 351-0198, Japan}
\affiliation{\mbox{Department of Applied Physics and Quantum-Phase Electronics Center, The University of Tokyo, Bunkyo-ku, Tokyo 113-8656, Japan}}
\affiliation{Tokyo College, The University of Tokyo, Bunkyo-ku, Tokyo 113-8656, Japan}

\author{Masashi~Tokunaga}
\affiliation{The Institute for Solid State Physics, The University of Tokyo, Kashiwa, Chiba 277-8581, Japan}

\author{Shingo~Yonezawa}
\email{yonezawa.shingo.3m@kyoto-u.ac.jp}
\affiliation{Department of Electrical, Electronic, and Digital Science and Engineering, Kyoto University, Kyoto, Kyoto 615-8510, Japan}

\date{\today}

\begin{abstract}
	First-order phase transitions often proceed through metastable states and complicated phase coexistence, yet such nonequilibrium processes are obscured in bulk measurements.
	The magneto-optical Kerr effect (MOKE) probes the local off-diagonal optical response, offering access to local magnetic dynamics beyond bulk measurements.
	Here, we investigate the magnetic-field-induced antiferromagnetic (AFM)--ferrimagnetic (FRM) transition in the polar insulator \FMO using pulsed-field MOKE measurements extended for the first time up to \SI{60}{\tesla} and down to \SI{1.4}{\kelvin}.
	The MOKE response exhibits unexpected and huge nonmonotonic transient behavior across the metamagnetic phase transition.
	The MOKE behavior does not show any scaling with the bulk magnetization, pointing to an MOKE analogue of the unconventional anomalous Hall effect.
	This is a new class of off-diagonal response emerging exclusively at the boundary between competing magnetic orders.
	At low temperatures, MOKE profiles are highly reproducible in detail, whereas at elevated temperatures they depend strongly on magnetic-field history and pulse protocol, suggesting changes in the driving mechanism of the phase transition.
	We infer that the excess MOKE anomaly originates from AFM--FRM phase interfaces in real space.
	Our results, extending MOKE into the \SIhyphen{60}{\tesla} regime, establish pulsed-field magneto-optics as a novel probe capable of uncovering transient and hidden field-induced phenomena through unconventional off-diagonal optical responses.
\end{abstract}

\makeatletter
\pretocmd{\titleblock@produce}{%
	\global\let\SMsavedAAC\@AAC@list
	\global\let\SMsavedAFF\@AFF@list
	\global\let\SMsavedAFG\@AFG@list
}{}{\PackageError{SMtitle}{Cannot snapshot REVTeX authors}{}}
\pretocmd{\frontmatter@footnote@produce}{%
	\global\let\SMsavedFMN\@FMN@list
}{}{\PackageError{SMtitle}{Cannot snapshot REVTeX notes}{}}
\makeatother
\maketitle
\raggedbottom


Off-diagonal electronic responses have been recognized as powerful probes of symmetry and topology in quantum materials~\cite{Xiao2010-us}.
In the dc limit, such responses are represented by the Hall and Nernst effects.
For example, the intrinsic anomalous Hall effect (AHE) reflects Berry curvature in momentum space~\cite{Yao2004-mn}, whereas noncoplanar spin textures can generate a topological Hall effect through real-space Berry phases~\cite{Neubauer2009-gq}.
Large anomalous Hall and Nernst responses in noncollinear antiferromagnets further demonstrate that off-diagonal transport can probe information that is not accessible by magnetization measurements~\cite{Nakatsuji2015-vo, Ikhlas2017-qw}.
At finite frequencies, off-diagonal optical conductivity gives rise to magneto-optical polarization rotation and circular dichroism~\cite{Feng2020-zp}.
Such optical off-diagonal responses are sensitive to magnetic symmetry and electronic structure~\cite{Higo2018-yn} and can be probed locally in both conductive and nonconductive materials~\cite{Huang2017-ns, Watanabe2026-sb}.

Among finite-frequency probes, the magneto-optical Kerr effect (MOKE), the polarization rotation of light reflected from magnetic samples, is a representative off-diagonal response~\cite{Qiu2000-wi}.
The Kerr angle $\thetaK$ originates from the optical Hall conductivity $\sigma_{xy}(\omega)$, while its magnitude also depends on the diagonal optical response $\sigma_{xx}(\omega)$~\cite{Feng2015-gv}.
Although conventionally used as a probe of magnetization, MOKE attracts renewed attention for its sensitivity to magnetic symmetry, spin--orbit coupling, and electronic structure, analogous to the dc Hall effect~\cite{Higo2018-yn, Iguchi2025-ce, Yoshimochi2026-nw}.
Its reflection geometry further enables measurements on optically opaque magnetic materials without requiring transmission through the sample.

\begin{figure}[t]
	\centering
	\includegraphics[width=246pt]{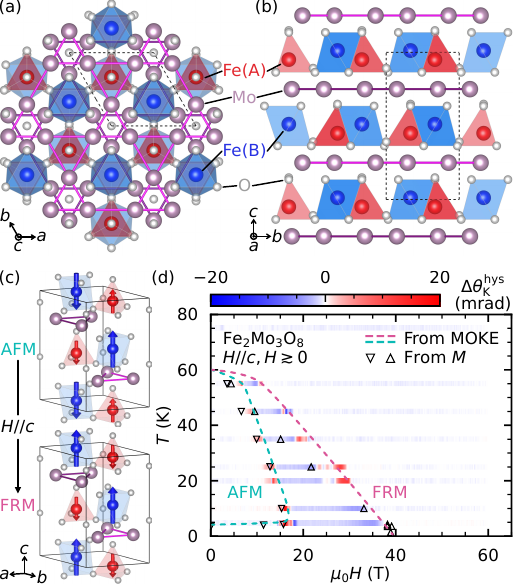}
	\caption{
		Crystal structure of \FMO viewed (a) along the $c$ axis and (b) perpendicular to it.
		Red and blue polyhedra show the oxygen coordination of the tetrahedral Fe(A) and octahedral Fe(B) sites, respectively; Mo and O are shown with purple and gray spheres.
		(c) Magnetic structures of the AFM and FRM states.
		Red and blue arrows indicate the Fe(A) and Fe(B) moments, respectively.
		(d) Field--temperature phase diagram for $\bm{H}\parallel c$.
		Triangles show the AFM--FRM transition fields determined from bulk magnetization~\cite{Chen2024-su}.
		Colors represent the MOKE hysteresis signal $\DTK^{\rm hys}$, defined as the difference between the down-sweep and up-sweep responses.
		The dashed curves are guides to the eye for the phase boundaries inferred from MOKE.
		The crystal structure was drawn using VESTA~\cite{Momma2011-oq}.
	}\label{fig:1}
\end{figure}

The polar insulator \FMO provides an intriguing playground in which magnetic symmetry, multiferroic behavior, and a field-driven metamagnetic phase transition are intertwined~\cite{Wang2015-gd, Kurumaji2015-qu}.
It crystallizes in the noncentrosymmetric polar space group $P6_3mc$ (No.~186).
Its crystal structure consists of Fe--O honeycomb layers containing tetrahedrally (A) and octahedrally (B) coordinated Fe$^{2+}$ sites, separated by nonmagnetic Mo--O sheets (Fig.~\ref{fig:1})~\cite{McCarroll1957-gs, Le-Page1982-ku, Kanazawa1986-xw}.
The Fe(A) sites have smaller magnetic moments than the Fe(B) sites~\cite{Varret1972-mc,Bertrand1975-wy}.
Below the \Neel temperature $\TN\simeq\SI{60}{\kelvin}$, the Fe moments form a collinear AFM state with the \Neel vector along the polar $c$ axis~\cite{Bertrand1975-wy, McAlister1983-yx}.
Within each Fe--O layer, the Fe(A) and Fe(B) moments order in an antiparallel configuration, whereas the total magnetic moment on each layer alternates along the $c$ axis, resulting in the AFM order with a zero net magnetization~\cite{Bertrand1975-wy, McAlister1983-yx, Wang2015-gd}.
This AFM state has recently been identified as a bi-altermagnet, in which the tetrahedral and octahedral Fe networks constitute distinct altermagnetic sublattices~\cite{Channagowdra2025-uh}.

In \FMO, a magnetic field applied along the $c$ axis drives a first-order metamagnetic transition from the AFM state to a ferrimagnetic (FRM) state with a finite net magnetization~\cite{McAlister1983-yx, Kurumaji2015-qu, Chen2024-su}.
A nonzero magnetic moment results from the parallel alignment of the layer magnetization, while the moments on Fe(A) and Fe(B) remain antiparallel to each other [Fig.~\ref{fig:1}(c)]~\cite{Wang2015-gd}.
The AFM--FRM transition fields determined from previous bulk magnetization measurements are plotted with triangles in Fig.~\ref{fig:1}(d)~\cite{Chen2024-su}.
This transition is strongly hysteretic, exhibiting pronounced field-history dependence in both magnetization and electric polarization~\cite{Chen2024-su}.
This large hysteresis makes \FMO a model system for investigating the nonequilibrium kinetics across a first-order phase transition.
Indeed, in Zn-substituted \FMO, real-space imaging studies resolve growth of AFM bubbles within an FRM matrix~\cite{Matsuura2023-ca}.
Analyses of the motion of the AFM--FRM phase interface have quantitatively linked thermally activated interface creep to the pronounced low-temperature hysteresis broadening~\cite{Matsuura2023-ca}.
Together with observations of AFM antiphase domain boundaries in undoped \FMO~\cite{Kim2018-lk}, these findings motivated us to investigate this metamagnetic transition and its dynamics using MOKE with an optical spot several micrometers in diameter.

Some of the present authors recently demonstrated pulsed-field MOKE measurements up to \SI{43}{\tesla}, establishing the feasibility of MOKE measurements under extreme magnetic fields~\cite{Ikeda2026-bq, Yamane2026-jl}.
In this work, we extend pulsed-field MOKE measurements up to \SI{60}{\tesla} and down to \SI{1.4}{\kelvin}.
We reveal gigantic MOKE responses that emerge at the field-driven first-order AFM--FRM phase boundary in \FMO.
The magnetic-field dependence of the MOKE response differs completely from that of the bulk magnetization, indicating the emergence of an unconventional anomalous MOKE response beyond magnetization scaling.
Its strong sweep-direction and pulse-protocol dependences point to nonequilibrium phase-transition dynamics.
We propose that the excess MOKE anomaly originates from moving AFM--FRM phase interfaces in real space.

We used the \FMO single crystal designated as K1 in Ref.~\cite{Chen2024-su}.
This crystal was grown by chemical vapor transport~\cite{Strobel1983-rd}.
Its crystallographic orientation was determined by X-ray Laue diffraction.
Basic properties of this sample are summarized in Fig.~\ref{app_fig:1}  in the Supplemental Material.
A nondestructive pulse magnet at the Institute for Solid State Physics, The University of Tokyo, was used to generate fields up to \SI{60}{\tesla} with a pulse duration of approximately \SI{36}{\milli\second}~\cite{Takeyama2006-us, Miyata2026-ur}.
The sample was cooled in a $^4$He cryostat as shown in Fig.~\ref{app_fig:1}(c).
The MOKE was measured in the polar geometry, which is sensitive to the out-of-plane magnetization, or, more generally, out-of-plane time-reversal symmetry breaking~\cite{Qiu2000-wi, Xia2006-rk, Kapitulnik2009-xh, Fried2014-uf, Schemm2014-cz, Higo2018-yn, Xu2022-hv, Yamane2026-bt}.
The Kerr angle was measured at a center wavelength of \SI{1550}{\nano\metre} using an all-fiber zero-area-loop Sagnac interferometer with a compact ferrule-based sample fixture~\cite{Xia2006-sg, Ikeda2026-bq, Yamane2026-bt}.
The interferometer selectively detects nonreciprocal phase shifts and is, in principle, insensitive to reciprocal linear birefringence~\cite{Xia2006-sg,Farhang2026-ab}.
Measurements were performed at a single optical spot on a surface normal to the $c$ axis, with $\bm{H}\parallel c\parallel-\bm{k}$ as illustrated in Fig.~\ref{app_fig:1}(b).
The spot diameter is approximately \SI{6}{\micro\metre}.
The positive field direction was defined to be along $-\bm{k}$, where $\bm{k}$ is the incident-light wave vector.
The optical arrangement and phase-resolved numerical lock-in analysis are described in detail in Refs.~\cite{Ikeda2026-bq,Yamane2026-jl, Yamane2026-rt}.
All MOKE data presented in the main text were acquired in positive-field pulses with an incident optical power of \SI{430}{\micro\watt} at the sample.
Results obtained under other conditions are presented in the Supplemental Material.

The measured polarization rotation angle contains a field-linear background, arising predominantly from the Faraday effect of optical components in the probe (see Supplemental Material).
For each measurement set, this background was estimated by a linear fit to the low-field AFM-state data from the lowest-temperature shot (see Fig.~\ref{app_fig:2} for an example).
The fitted linear component was then subtracted from all data in the same measurement set to obtain the background-subtracted Kerr angle $\DTK$.

\begin{figure}[t]
	\centering
	\includegraphics[width=246pt]{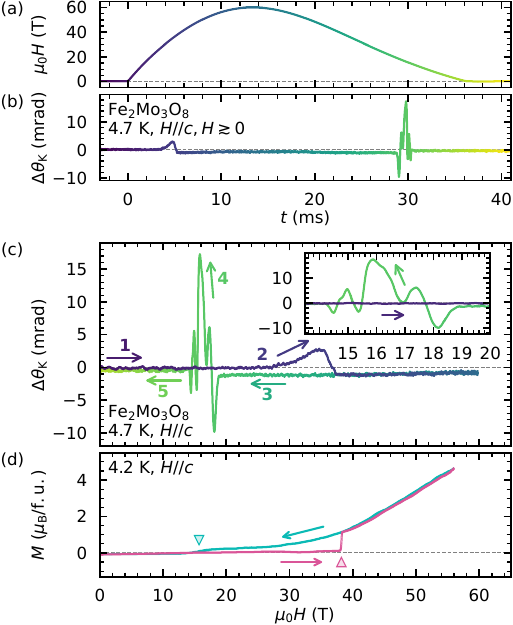}
	\caption{
		Pulsed-field MOKE response of \FMO at \SI{4.7}{\kelvin}. (a) Magnetic field and (b) Kerr angle versus time. (c) Same MOKE data plotted as a function of magnetic field.
		Colors in (a)--(c) progressing from purple to yellow indicate time evolution.
		The numbered arrows in (c) show the sequence along the pulse.
		The inset in (c) shows an enlarged view of the structured down-sweep data.
		(d) Bulk magnetization measured previously at \SI{4.2}{\kelvin} on the same crystal~\cite{Chen2024-su}.
		Magenta and cyan denote the up- and down-sweeps, respectively, with arrows indicating the sweep directions.
		Open up- and down-pointing triangles mark the corresponding bulk metamagnetic transition fields.
	}\label{fig:2}
\end{figure}

Figure~\ref{fig:2} shows the pulsed-field MOKE response of \FMO at \SI{4.7}{\kelvin}.
The magnetic-field pulse reaches \SI{60}{\tesla} [Fig.~\ref{fig:2}(a)], and the MOKE response is successfully observed throughout the pulse.
During the pulse, $\DTK$ exhibits a pronounced nonmonotonic response [Fig.~\ref{fig:2}(b)].
The field dependence of MOKE differs markedly between the up- and down-sweeps [Fig.~\ref{fig:2}(c)].
On the up-sweep, $\DTK$ develops a broad positive hump centered near \SI{35}{\tesla}, reaching approximately \SI[retain-explicit-plus]{+3}{\milli\radian}, and then shows a change to a small negative value in the FRM state.
Notice that $\DTK$ in the FRM state, $\sim\SI{-1}{\milli\radian}$, differs from $\DTK\sim\SI{0}{\milli\radian}$ in the AFM phase.
This difference is attributable to the finite net magnetic moment in the FRM phase~\cite{Chen2024-su}.
The down-sweep data exhibit a much larger and more structured spike anomaly in the field range of \SIrangeDash{14}{19}{\tesla}.
The MOKE response therefore displays a pronounced sweep-direction dependence, not only in its transition fields but also in its amplitude, lineshape, and polarity.
We comment that the sample reflectivity exhibits only minute variation across the transition, indicating that the observed MOKE anomaly cannot be attributed to spectral changes (see Supplemental Material).

For comparison, Fig.~\ref{fig:2}(d) shows the bulk magnetization measured previously at \SI{4.2}{\kelvin} on the same crystal~\cite{Chen2024-su}.
The MOKE anomalies occur within the field range of the AFM--FRM transition identified from the magnetization data [marked by triangles in Fig.~\ref{fig:2}(d)].
This correspondence indicates that the MOKE anomalies originate from the first-order metamagnetic phase transition.
We also confirmed that these anomalies are absent in field sweeps that do not reach the FRM phase (see Fig.~\ref{app_fig:4} in the Supplemental Material).
Remarkably, the MOKE and magnetization curves exhibit qualitatively different lineshapes: the bulk magnetization shows a sharp step-like change on the up-sweep and a broader step on the down-sweep, whereas $\DTK$ exhibits a broad hump and a pronounced sign change, respectively.
Thus, although the MOKE anomalies track the metamagnetic phase boundary, they do not show conventional scaling with the bulk magnetization.
Moreover, the up-sweep MOKE data exhibit a precursor-like hump within the AFM phase before the metamagnetic transition, with no corresponding feature in the bulk magnetization.
Because this response does not follow conventional magnetization scaling, we refer to it as an unconventional anomalous MOKE response, an optical analogue of the unconventional anomalous Hall effect~\cite{Nagaosa2010-qh, Taguchi2001-rg, Kanazawa2011-ff, Ishizuka2018-kd, Suzuki2016-eu, Feng2020-zp, Kato2023-fp}.

\begin{figure}[t]
	\centering
	\includegraphics[width=246pt]{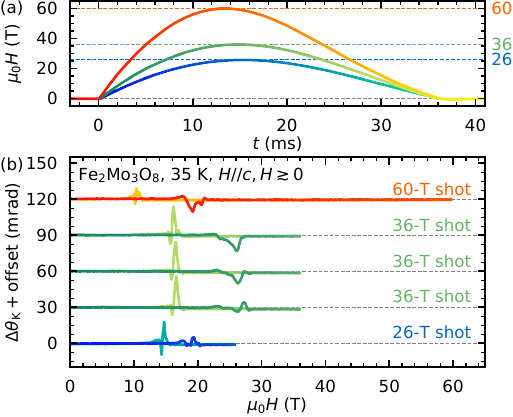}
	\caption{
		Pulse-protocol dependence at \SI{35}{\kelvin}, showing thermal-fluctuation-driven phase transition dynamics.
		(a) Field waveforms and (b) corresponding $\DTK(H)$ data for pulses with peak fields of \SI{26}{\tesla}, \SI{36}{\tesla}, and \SI{60}{\tesla}.
		The three \SIhyphen{36}{\tesla} datasets were obtained in separate shots.
		Color gradients encode time within each pulse; the blue, green, and red traces correspond to \SI{26}{\tesla}, \SI{36}{\tesla}, and \SI{60}{\tesla}, respectively.
		Dashed horizontal lines in (a) mark the nominal peak fields and zero field.
		$\DTK$ data in (b) are shifted upward by \SI{30}{\milli\radian}, with dashed lines marking the corresponding zero levels.
	}\label{fig:3}
\end{figure}

Figure~\ref{fig:3} compares the MOKE responses measured under different field pulses with peak heights of \SI{26}{\tesla}, \SI{36}{\tesla}, and \SI{60}{\tesla} at \SI{35}{\kelvin}.
Note that the three \SIhyphen{36}{\tesla} pulse datasets were obtained in separate shots, yet their field waveforms are almost identical [Fig.~\ref{fig:3}(a)].
Surprisingly, the position and lineshape of the MOKE anomalies vary markedly among the different field pulses.
Moreover, even among the three \SIhyphen{36}{\tesla} pulses, the MOKE response varies substantially between shots.
These observations indicate that the MOKE responses at \SI{35}{\kelvin} are not uniquely determined by the instantaneous magnetic field.
In contrast, the low-temperature MOKE responses are highly reproducible between repeated field pulses (Fig.~\ref{app_fig:3} in the Supplemental Material).
This temperature dependence of the reproducibility suggests that the AFM--FRM transition process becomes less susceptible to thermal fluctuations at low temperature, whereas thermally activated processes dominate at elevated temperatures.

\begin{figure}[t]
	\centering
	\includegraphics[width=246pt]{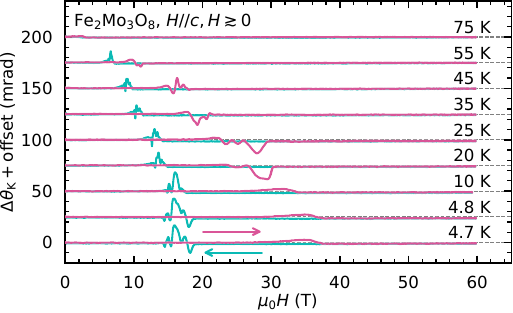}
	\caption{
		Temperature evolution of $\DTK(H)$ measured with \SIhyphen{60}{\tesla} pulses.
		Magenta and cyan curves denote the up- and down-sweeps, respectively, and arrows indicate the sweep directions.
		$\DTK$ data are shifted upward by \SI{25}{\milli\radian} for clarity, with dashed lines marking the corresponding zero levels.
	}\label{fig:4}
\end{figure}

Given the substantial pulse-profile dependence, we examine the temperature evolution of the MOKE response using a common \SIhyphen{60}{\tesla} pulse profile (Fig.~\ref{fig:4}).
At \SI{4.7}{\kelvin}, \SI{4.8}{\kelvin}, and \SI{10}{\kelvin}, the up-sweep response exhibits similar broad positive humps.
On warming, these humps evolve into sign-changing features.
The down-sweep spikes also gradually decrease in amplitude and become less structured with increasing temperature.
The anomalies are no longer clearly resolved at \SI{75}{\kelvin}.

The hysteretic MOKE response is summarized in the $H$--$T$ diagram in Fig.~\ref{fig:1}(d), where the colors represent the branch difference $\DTK^{\rm hys}(H,T)=\DTK^{\rm down}(H,T)-\DTK^{\rm up}(H,T)$.
This map highlights the hysteretic region associated with the AFM--FRM transition.
The dashed curves serve as guides to the eye for the MOKE-derived phase boundaries.
The characteristic MOKE anomaly fields similarly shift to lower fields on warming and follow the overall temperature dependence of the AFM--FRM transition fields determined from bulk magnetization~\cite{Chen2024-su}, although the two sets of fields do not coincide exactly.
These differences in the phase boundaries may reflect the local nature of the MOKE measurement, with a spot diameter of approximately \SI{6}{\micro\metre}, or small differences in the magnetic-field pulse profiles.
At \SI{1.4}{\kelvin}, it is known that the hysteresis becomes so wide that the field-induced FRM phase persists down to zero field~\cite{Chen2024-su}.
Our MOKE data at \SI{1.4}{\kelvin} likewise reveal an absence of the spikes in the down-sweep branch and nonzero $\DTK$ even at $H=0$ (Fig.~\ref{app_fig:5} in the Supplemental Material).
These results are consistent with the reported hysteresis broadening~\cite{Chen2024-su}.

\begin{figure}[t]
	\centering
	\includegraphics[width=246pt]{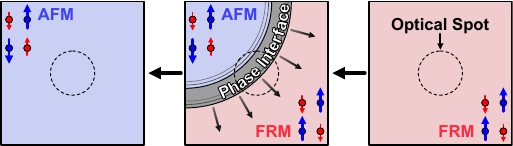}
	\caption{
		Schematic illustration of the proposed interfacial origin of the down-sweep MOKE transient.
		The \SIhyphen{6}{\micro\metre} optical spot is indicated by the dashed circle.
		The right panel shows the real-space distribution of the FRM phase at higher field.
		As the field decreases, AFM regions nucleate and then the FRM phase fraction decreases as the AFM--FRM phase interface moves (middle panel).
		If the interface has a magneto-optical response distinct from those of the bulk phases, its passage through the optical spot can produce a pronounced transient.
		The left panel illustrates recovery of the AFM state at low field.
	}\label{fig:5}
\end{figure}

We now discuss possible origins of the spike-like MOKE features and the low-temperature up-sweep hump.
The spike-like MOKE response cannot be explained by any weighted sum of optical responses of the AFM and FRM bulk phases.
This indicates that novel metastable structures with excess MOKE responses are formed at the metamagnetic phase boundary.
We propose that these structures are real-space AFM--FRM phase interfaces (Fig.~\ref{fig:5}).
Indeed, AFM--FRM phase interfaces have been observed in MOKE and magnetic-force-microscopy (MFM) imaging studies on Zn-substituted \FMO~\cite{Matsuura2023-ca}.
In Ref.~\cite{Matsuura2023-ca}, AFM bubbles are observed, and they are found to grow within an FRM matrix in the metastable FRM state prepared by decreasing the field.
Moreover, the MOKE images suggest that the AFM--FRM interfaces exhibit MOKE responses distinct from those in the AFM or FRM phases, whereas the MFM images do not show an enhancement of magnetization at the interface.
In this picture, as the field decreases across the metamagnetic transition, AFM regions nucleate, and their growth causes rapid movement of the phase interfaces.
Their passage through the optical spot would produce the spike-like MOKE responses.

The microscopic magnetic structure of the proposed interface remains an open question.
One possibility is a metastable spin-flopped state with in-plane \Neel-vector components.
Spin flop in this material is energetically disfavored by
the strong uniaxial anisotropy~\cite{Ghara2023-rg}, but may be transiently realized in the interface.
Such a spin-flopped magnetic configuration in the bi-altermagnet \FMO resembles the magnetic structure of the altermagnet \ce{MnTe}, which is known to exhibit a large spontaneous MOKE response~\cite{Watanabe2026-sb}.
Another possibility is a long-period noncollinear spin texture at the AFM--FRM interface, analogous to the long-range spin textures accompanying ferromagnetic domain walls.
If such an interfacial spin texture is noncoplanar, it could generate an emergent magnetic field through scalar spin chirality, potentially contributing to the off-diagonal optical response~\cite{Feng2020-zp,Kato2023-fp,Farhang2026-ab,Shimizu2026-rq}.

The hump in the low-temperature up-sweeps should have a different mechanism, since it occurs within the AFM phase.
We interpret the broad hump as a precursor MOKE response associated with magnetic fluctuations or weak spin rotation preceding the metamagnetic transition.
It is also interesting that the spike-like MOKE feature is absent when this hump is observed.
This indicates that the transition dynamics differ from those associated with partial phase nucleation followed by phase-interface creep, as discussed above.
The absence of spikes suggests a more abrupt and homogeneous phase transition.
This is supported by the fact that the metamagnetic transition in magnetization is very sharp only in the low-temperature up-sweeps, whereas the transition is much broader under other conditions~\cite{Chen2024-su}.
The high reproducibility of the MOKE response at low temperature suggests a minor thermal contribution.
Thus, the homogeneous phase transition at low temperatures should be driven by quantum fluctuations, which have perhaps been detected as the MOKE humps.

In summary, by extending pulsed-field MOKE to the \SIhyphen{60}{\tesla} regime and down to \SI{1.4}{\kelvin}, we have discovered unconventional off-diagonal optical responses across the first-order metamagnetic transition in \FMO.
This discovery highlights boundaries between competing magnetic phases as a promising realm for exploring and potentially engineering unconventional phenomena related to topology and broken symmetry.
Our observations demonstrate that pulsed-field MOKE is sensitive to local and nonequilibrium magnetic features that are not captured by the volume-averaged bulk magnetization.
Thus, pulsed-field MOKE measurements in extreme conditions will be a powerful new probe of magnetic-field-induced phenomena including field-driven phase transitions.

\section*{Acknowledgements}
	We acknowledge I.~Kakeya and Y.~Gotoh for valuable comments and discussions.
	We also thank A.~Kirikoshi and J.~Otsuki for theoretical discussions.
	This work was supported by Grant-in-Aids for Scientific Research (KAKENHI Grant Nos. JP23K17670, JP23H04861, JP23H04862, JP24H00194, JP24H01663, JP25K17341, JP26H00588, JP26H00656, and JP26K17095) from the Japan Society for the Promotion of Science,
	by ISHIZUE 2023 of Kyoto University Research Development Program,
	by Iketani Science and Technology Foundation (Grant No.~0361078-A),
	and by The Mitsubishi Foundation (Grant No.~202410051).
	This work was carried out under the Visiting Researcher's Program of the Institute for Solid State Physics, the University of Tokyo (ISSPkyodo-202511-HMBXX-0020).
	We acknowledge support for the construction of experimental setups from Research Equipment Development Support Room of the Graduate School of Science, Kyoto University; and support for liquid helium and nitrogen supplies from Low Temperature and Materials Sciences Division, Agency for Health, Safety and Environment, Kyoto University.

	\bibliographystyle{apsrev4-2-arxivdot}
	\bibliography{paperpile}

	\makeatletter
	\def\hyper@natlinkstart#1{%
		\Hy@backout{#1}%
		\hyper@linkstart{cite}{cite.SM.#1}%
		\def\hyper@nat@current{#1}%
	}
	\def\hyper@natlinkbreak#1#2{%
		\hyper@linkend#1\hyper@linkstart{cite}{cite.SM.#2}%
	}
	\newcommand\SMmarkcitation[1]{%
		\expandafter\gdef\csname SM@cited@#1\endcsname{1}%
	}
	\pretocmd{\@citex}{\forcsvlist{\SMmarkcitation}{#3}}{}%
	{\PackageError{SMrefs}{Could not track SM citations}{}}
	\makeatother

	\clearpage
	\setcounter{secnumdepth}{2}

	\setcounter{section}{0}
	\setcounter{figure}{0}
	\setcounter{table}{0}
	\setcounter{equation}{0}

	\renewcommand{\thesection}{S\arabic{section}}
	\renewcommand{\thesubsection}{\thesection.\arabic{subsection}}
	\renewcommand{\thefigure}{S\arabic{figure}}
	\renewcommand{\theHfigure}{S\arabic{figure}}
	\renewcommand{\thetable}{S\arabic{table}}
	\renewcommand{\theequation}{S\arabic{equation}}

	\newlength{\SMparindent}
	\setlength{\SMparindent}{\parindent}
	\newlength{\SMparskip}
	\setlength{\SMparskip}{\parskip}
	\newcommand{\SMtextparagraphs}{%
		\setlength{\parindent}{\SMparindent}%
		\setlength{\parskip}{\SMparskip}%
	}
	\renewcommand{\thepage}{S\arabic{page}}
	\setcounter{page}{1}

	\setcounter{figure}{0}
	\renewcommand{\thefigure}{S\arabic{figure}}

	\setcounter{table}{0}
	\renewcommand{\thetable}{S\arabic{table}}

	\setcounter{equation}{0}
	\renewcommand{\theequation}{S\arabic{equation}}

	\onecolumngrid
	\begin{center}
		{\large\bfseries Supplemental Material for\\[0.3em]
			``\maintitle''}
	\end{center}
	\makeatletter
	\begingroup
	\frontmatter@setup
	\def\@mpfn{mpfootnote}
	\let\@makefnmark\frontmatter@makefnmark
	\let\@AAC@list\SMsavedAAC
	\let\@AFF@list\SMsavedAFF
	\let\@AFG@list\SMsavedAFG
	\let\@FMN@list\SMsavedFMN
	\def\frontmatter@footnotemark#1{%
		\begingroup
		\csname c@\@mpfn\endcsname#1\relax
		\def\@thefnmark{\frontmatter@thefootnote}\@makefnmark
		\endgroup
	}
	\expandafter\def\csname thefootnote\endcsname{\fnsymbol{footnote}}
	\def\SMprintnotes{%
		\def\@TBN@opr##1##2{%
			\footnotetext[##1]{\csname textsuperscript\endcsname{\@fnsymbol{##1}}~##2}%
		}%
		\@FMN@list
	}
	\let\frontmatter@footnote@produce\SMprintnotes
	\groupauthors@sw{\frontmatter@author@produce@group}{\frontmatter@author@produce@script}
	\endgroup
	\makeatother
	\vspace{0.6\baselineskip}

	\par\noindent\begin{minipage}[b]{510pt}
		\makeatletter\def\@captype{figure}\makeatother
		\centering
		\includegraphics[width=510pt]{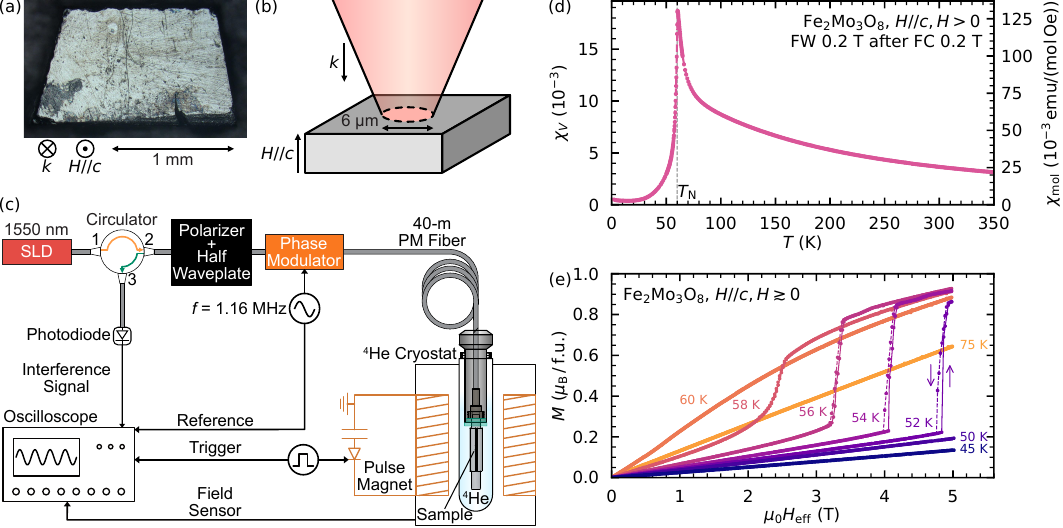}
		\caption{
			Sample characterization of \FMO and optical setup for pulsed-field MOKE measurements.
			(a) Photograph of the \FMO single crystal K1.
			(b) Polar MOKE geometry with $\bm{H}\parallel c\parallel-\bm{k}$ and an optical spot size of approximately \SI{6}{\micro\metre}.
			(c) All-fiber zero-area-loop Sagnac interferometer for pulsed-field MOKE measurements at \SI{1550}{\nano\metre}.
			(d) Volume susceptibility $\chi_V$ in SI units (left axis) and molar susceptibility $\chi_{\mathrm{mol}}$ in cgs units (right axis) of the \FMO K1 sample measured with a SQUID magnetometer on warming in \SI{0.2}{\tesla} after field cooling in the same field.
			The vertical dashed line marks $\TN\simeq\SI{60}{\kelvin}$.
			(e) Magnetic moment per formula unit $M$ versus effective magnetic field $\mu_0 H_{\rm eff}$ for $\bm{H}\parallel c$ measured at the indicated temperatures.
			Solid and dashed curves denote increasing and decreasing fields, respectively.
			The magnetometry data in (d) and (e) are corrected by taking into account the demagnetization effect.
		}\label{app_fig:1}
	\end{minipage}\par
	\twocolumngrid

\section{Sample characterization and experimental setup}

	The measurements were performed on the \FMO single crystal K1, which was also used in the previous bulk magnetization study of Ref.~\cite{Chen2024-su}.
	The crystal was grown by chemical vapor transport~\cite{Strobel1983-rd}, and its orientation was determined by X-ray Laue diffraction.
	Figure~\ref{app_fig:1}(d) shows the magnetic susceptibility measured with a SQUID magnetometer on warming in \SI{0.2}{\tesla} after field cooling in the same field.
	The sharp anomaly in the magnetic susceptibility indicates AFM ordering at $\TN\simeq\SI{60}{\kelvin}$ in agreement with previous reports.
	The $M$--$\mu_0 H_{\rm eff}$ curves for $\bm{H}\parallel c$ exhibit metamagnetic transitions and hysteresis at temperatures below $\TN$ within the measured field range [Fig.~\ref{app_fig:1}(e)].
	These data are consistent with the previous reports~\cite{McAlister1983-yx, Wang2015-gd, Kurumaji2015-qu}.

	Polar MOKE was measured at \SI{1550}{\nano\metre} using a broadband superluminescent diode (SLD) and an all-fiber zero-area-loop (loop-less) Sagnac interferometer [Fig.~\ref{app_fig:1}(c)]~\cite{Xia2006-sg, Ikeda2026-bq, Yamane2026-jl, Yamane2026-bt}.
	\clearpage
	\onecolumngrid

	\noindent\begin{minipage}[t]{\dimexpr(\textwidth-\columnsep)/2\relax}
		\vspace{0pt}
		\par\noindent\begin{minipage}[b]{\linewidth}
			\makeatletter\def\@captype{figure}\makeatother
			\centering
			\includegraphics[width=246pt]{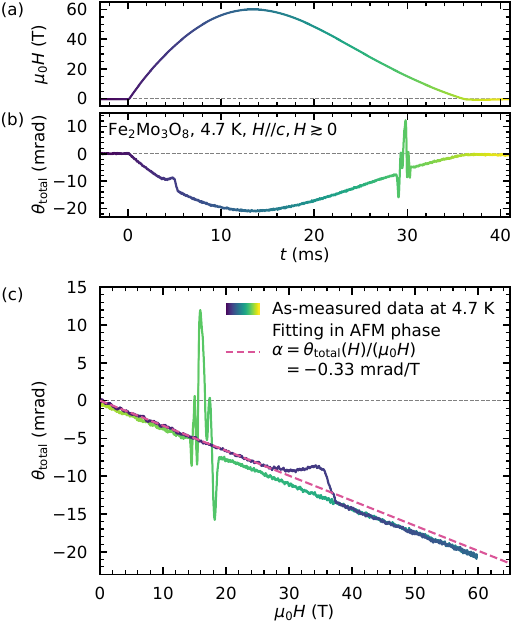}
			\caption{
				Subtraction of the field-linear component in pulsed-field MOKE measurements.
				(a) Magnetic field and (b) total magneto-optical angle $\thetaT$ versus time at \SI{4.7}{\kelvin}.
				Colors in (a)--(c) progress from purple to yellow with time.
				(c) As-measured $\thetaT$ as a function of magnetic field.
				The magenta dashed line shows a field-linear fit to the AFM-state data, with slope $\alpha = \SI{-0.33}{\milli\radian\per\tesla}$.
				The background-subtracted Kerr angle is defined as $\DTK=\thetaT-\alpha\mu_0H$.
				The same fitted linear component is subtracted from all shots acquired during the same experimental set.
			}\label{app_fig:2}
		\end{minipage}\par\medskip
		\par\noindent\begin{minipage}[b]{\linewidth}
			\makeatletter\def\@captype{figure}\makeatother
			\centering
			\includegraphics[width=246pt]{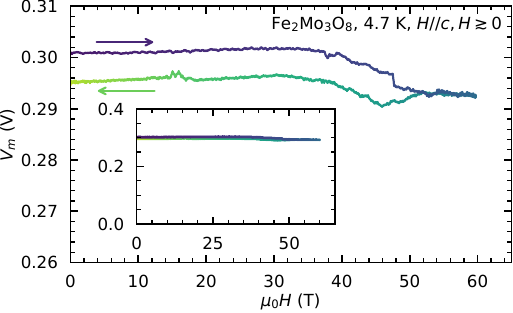}
			\caption{
				Field dependence of the interference amplitude $V_m$ during a pulsed-field MOKE measurement at \SI{4.7}{\kelvin} for $H\parallel c$.
				Here, $V_m$ is proportional to the optical coupling to the sample and sample reflectivity.
				$V_m$ varies by approximately \SI{4}{\percent} peak to peak during the pulse.
				The inset shows the same data on a vertical scale starting at zero.
			}\label{app_fig:2p5}
		\end{minipage}\par
	\end{minipage}\hfill
	\begin{minipage}[t]{\dimexpr(\textwidth-\columnsep)/2\relax}
		\vspace{0pt}
		\SMtextparagraphs
		The incident light was focused to an approximately \SIhyphen{6}{\micro\metre} spot on a surface normal to the $c$ axis.
		Thus, the MOKE measurements probe magnetic properties within this illuminated region.
		The magnetic field was applied parallel to the $c$ axis so that the polar MOKE geometry was realized with $\bm{H}\parallel c\parallel-\bm{k}$ [Fig.~\ref{app_fig:1}(b)].
		The positive field direction is defined to be along $-\bm{k}$.
		The setup uses a \SIhyphen{40}{\metre} polarization-maintaining (PM) fiber and phase modulation at \SI{1.16}{\mega\hertz}.
		The modulation depth $\phi_m$ was set to about 2.1 rad, close to the value that maximizes $J_3 (2\phi_m)$, where $J_3$ is the third-order Bessel function of the first kind~\cite{Ikeda2026-bq}.
		The as-measured angle $\thetaT$ was extracted from the first through sixth harmonic voltages using the equation in Ref.~\cite{Ikeda2026-bq}.
		The photodiode signal was recorded together with the waveform of the modulation drive voltage and a field-sensor (pickup-coil) signal.
		A nondestructive pulse magnet generated fields up to \SI{60}{\tesla} with a pulse duration of approximately \SI{36}{\milli\second}, and the sample was cooled in a $^4$He cryostat~\cite{Takeyama2006-us, Miyata2026-ur}.

		\section{Subtraction of the field-linear component from the magneto-optical angle}

			The total angle $\thetaT$ of the measured magneto-optical effect was extracted using phase-resolved numerical lock-in analysis described in Refs.~\cite{Ikeda2026-bq,Yamane2026-jl, Yamane2026-rt}.
			We note that $\thetaT$ contains a substantial field-linear contribution arising predominantly from the Faraday effect of optical components in the probe.
			We find that the field-linear contribution varies each time the optical probes are reinserted into the cryostat [Fig.~\ref{app_fig:1}(c)].
			Therefore, we evaluate this contribution for each set of measurements.
			We performed linear fitting to the pure AFM-state data of the lowest-temperature shot.
			The fitted linear component $\alpha \mu_0H$ was then subtracted from all records obtained in the same set of measurements to obtain the Kerr angle, $\DTK(H)=\thetaT(H)-\alpha \mu_0H$.

			Figure~\ref{app_fig:2} illustrates this procedure for a measurement at \SI{4.7}{\kelvin}, for which the fitted slope $\alpha$ with respect to $\mu_0H$ yields \SI{-0.33}{\milli\radian\per\tesla}.
			Across all measurement sets, the fitted slope $\alpha$ ranges from \SI{-0.38}{\milli\radian\per\tesla} to \SI{-0.13}{\milli\radian\per\tesla}.
			We comment that this linear component may contain a small contribution intrinsic to the sample.
			Accordingly, $\DTK$ should be interpreted as the response relative to this subtracted component rather than as the absolute Kerr angle of the sample.

		\section{Optical coupling and reflection from the sample surface}
			Figure~\ref{app_fig:2p5} shows the field dependence of $V_m$ at \SI{4.7}{\kelvin}.
			Here, $V_m$ is the amplitude of the interference signal and depends on the optical coupling to the sample and the sample reflectivity~\cite{Ikeda2026-bq}.
			Assuming constant optical coupling, the observed variation in $V_m$ indicates that the sample reflectivity varies by approximately \SI{4}{\percent} peak to peak during the pulse.

	\end{minipage}\par
	\clearpage

	\par\noindent\begin{minipage}[b]{510pt}
		\makeatletter\def\@captype{figure}\makeatother
		\centering
		\includegraphics[width=510pt]{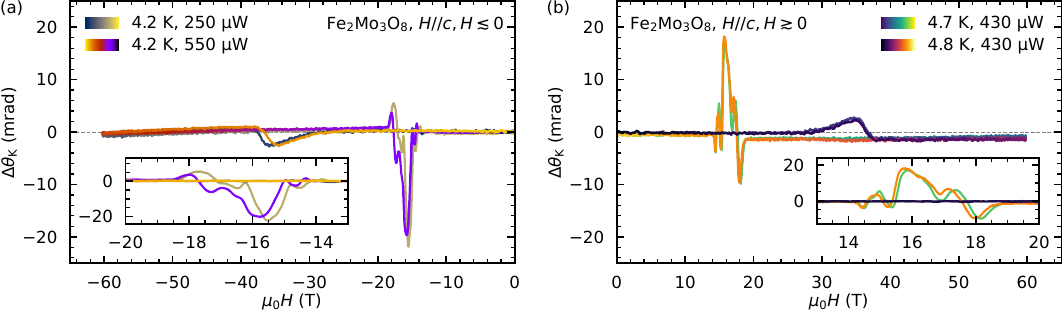}
		\caption{
			Low-temperature MOKE data of \FMO measured in (a) negative and (b) positive \SIhyphen{60}{\tesla} field pulses.
			Temperatures and optical power are indicated in the legends, and color gradients encode time evolution within each pulse.
			The insets show enlarged views of the unconventional return-branch transients.
			The small shifts of the anomaly fields in (a) are consistent with minute local heating caused by higher optical power.
			The four datasets exhibit very similar shapes despite the small temperature differences and different incident optical power.
		}\label{app_fig:3}
	\end{minipage}\par
	\twocolumngrid
	For example, a possible spectral shift of crystal-field $d$--$d$ transitions at the AFM--FRM interface could affect the reflectivity at \SI{0.8}{\electronvolt} (\SI{1550}{\nano\meter}).
	The small variation in $V_m$ disfavors spectral changes that would produce a large change in the measured reflectivity at this wavelength.

\section{Low-temperature reproducibility and optical-power dependence}

	Figure~\ref{app_fig:3} compares additional low-temperature data measured in negative and positive field pulses.
	The positive-field data were acquired with an incident optical power of \SI{430}{\micro\watt} at the sample.

	The two datasets exhibit very similar transient profiles, including detailed structures in the return-branch anomaly [Fig.~\ref{app_fig:3}(b)].
	This agreement shows that the low-temperature MOKE curves are highly reproducible.

	For the negative-field measurements at \SI{4.2}{\kelvin} [Fig.~\ref{app_fig:3}(a)] with incident optical power of \SI{250}{\micro\watt} and \SI{550}{\micro\watt}, MOKE signals exhibit almost the same behavior as the positive-field measurements, apart from the sign change [Fig.~\ref{app_fig:3}(b)].
	We nevertheless observed small shifts of the anomaly fields in the higher-power data; the metamagnetic hysteresis becomes slightly narrower.
	These changes can be attributed to local heating, since a small temperature rise in this temperature regime is known to narrow the hysteresis~\cite{Chen2024-su}.
	Importantly, the possible heating effect is rather minute, showing that the data in this paper reflect the intrinsic properties of \FMO.

\section{MOKE response with and without the AFM--FRM transition}

\begin{figure}[tbp]
		\centering
		\includegraphics[width=246pt]{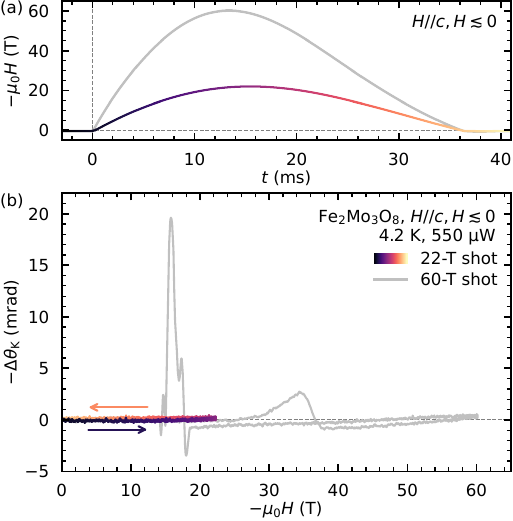}
		\caption{
			Negative-field MOKE measurements of \FMO at \SI{4.2}{\kelvin} with an incident optical power of \SI{550}{\micro\watt}.
			(a) Magnetic-field waveforms for pulses with peak magnitudes of \SI{22}{\tesla} (colored) and \SI{60}{\tesla} (gray).
			(b) Kerr angle plotted as $-\DTK$ versus $-\mu_0H$.
			Notice that the signs of horizontal and vertical axes are inverted to provide better comparison with the positive-pulse data.
			The color gradient shows time evolution within the \SIhyphen{22}{\tesla} pulse, and arrows indicate field change directions.
			These data demonstrate that the peak in MOKE is absent if the field does not reach the up-sweep metamagnetic transition field (approximately \SI{38}{\tesla} at \SI{4.2}{\kelvin}).
		}\label{app_fig:4}
	\end{figure}

	We measured the MOKE response in a negative-field pulse with a peak magnitude of \SI{22}{\tesla} at \SI{4.2}{\kelvin}, below the AFM--FRM transition field for increasing $|H|$.
	Figure~\ref{app_fig:4} compares the \SIhyphen{22}{\tesla} and \SIhyphen{60}{\tesla} shot data with an incident optical power of \SI{550}{\micro\watt}.

	In the \SIhyphen{22}{\tesla} shot, where the sample stays in the AFM phase, $\DTK$ remains close to zero on both sweeps.
	No comparable anomaly is observed at field magnitudes of \SIrangeDash{14}{19}{\tesla}, where the \SIhyphen{60}{\tesla} data show large spikes as $|H|$ decreases.
	This comparison supports the association of the MOKE anomaly with the AFM--FRM transition.

\section{Successive negative-field pulses at 1.4 K}

	\begin{figure}[tbp]
		\setlength{\abovecaptionskip}{3pt}
		\centering
		\includegraphics[width=246pt]{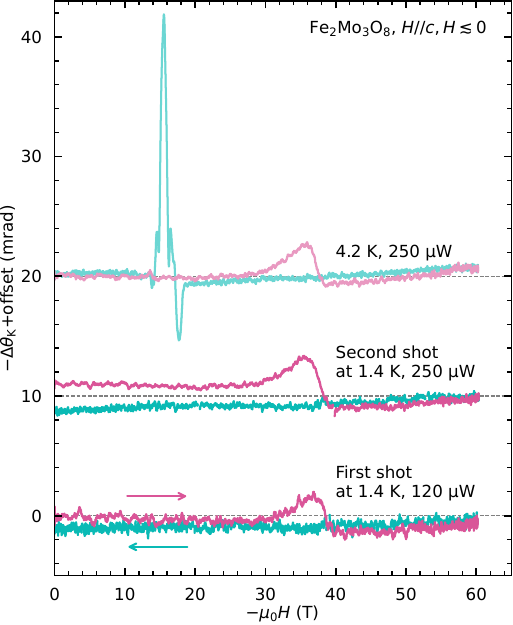}
		\caption{
			Successive negative-field MOKE results at \SI{1.4}{\kelvin}, acquired with optical power of \SI{120}{\micro\watt} for the first shot and \SI{250}{\micro\watt} for the second shot.
			The upper pair of curves shows a measurement at \SI{4.2}{\kelvin} with \SI{250}{\micro\watt} for comparison.
			Signals are plotted as $-\DTK$ versus $-\mu_0H$ to provide better comparison with the positive-pulse data.
			Magenta and cyan curves denote data under increasing and decreasing $|H|$, respectively, and arrows indicate the sweep directions.
			Successive records are shifted upward by \SI{10}{\milli\radian}, and dashed lines mark the corresponding zero levels.
			Both \SIhyphen{1.4}{\kelvin} shots show an anomaly on the outward branch, but no reverse-transition transient is resolved on the return branch.
			The zero-field values differ between the two branches.
			The reduced signal-to-noise ratio in the first shot reflects the low optical power and liquid-helium boiling.
		}\label{app_fig:5}
	\end{figure}

	\begingroup
	\widowpenalty=10000
	Figure~\ref{app_fig:5} shows MOKE data at \SI{1.4}{\kelvin} from two successive negative-field shots, acquired with optical power of \SI{120}{\micro\watt} and \SI{250}{\micro\watt}, respectively.
	In the first shot, a hump anomaly develops as $|H|$ increases, whereas no reverse-transition transient is observed as $|H|$ decreases.
	A nonzero MOKE signal remains at zero field after the pulse.
	We attribute this behavior to the persistence of the field-induced FRM state after removal of the magnetic field.
	This widening of the hysteresis region down to zero field is consistent with the previous reports~\cite{Chen2024-su, Matsuura2023-ca}.
	\par
	\endgroup

	\begingroup
	\interlinepenalty=10000
	The second shot starts from a nonzero Kerr angle and again exhibits an outward-branch anomaly, followed by a remanent MOKE signal.
	Between the first and second shots, the sample temperature was kept at \SI{1.4}{\kelvin} for about \SI{55}{\minute}.
	We interpret the nonzero initial Kerr angle together with the renewed transition anomaly as evidence for coexisting AFM and FRM regions before the second pulse.
	The magnetic state in part of the sample may relax into the AFM phase very slowly between the pulses.
	Within this picture, the outward-branch anomaly reflects the transition of the remaining AFM regions into the FRM state.
	\par
	\endgroup

	For comparison, the upper curves in Fig.~\ref{app_fig:5} show data measured at \SI{4.2}{\kelvin} with an optical power of \SI{250}{\micro\watt}.
	A pronounced transient is observed on the return branch as discussed in the main text, unlike in the two \SIhyphen{1.4}{\kelvin} shots.

\section{Temperature evolution near the magnetic ordering temperature}

	Figure~\ref{app_fig:6} presents measurements in pulsed fields of approximately \SI{15}{\tesla} near $\TN$.
	The anomalies observed at \SIrangeDash{53}{57}{\kelvin} weaken on warming and are no longer clearly observed at \SI{59}{\kelvin} and above.
	This suppression of the metamagnetism near the ordering temperature is consistent with the bulk susceptibility [Fig.~\ref{app_fig:1}(d)].

	\par\noindent\begin{minipage}[b]{\linewidth}
		\makeatletter\def\@captype{figure}\makeatother
		\centering
		\includegraphics[width=246pt]{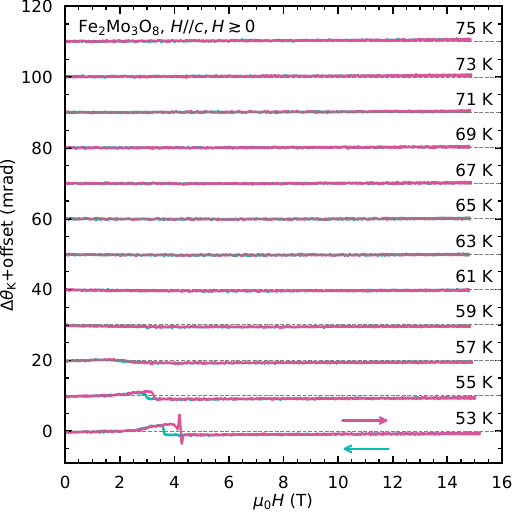}
		\caption{
			Temperature evolution of $\DTK$ in pulsed fields of approximately \SI{15}{\tesla} near $\TN$.
			Magenta and cyan arrows indicate the up- and down-sweep directions, respectively.
			Successive records are offset by \SI{10}{\milli\radian}, and dashed lines mark the corresponding zero levels.
			The anomalies weaken on warming and are no longer clearly resolved at \SI{59}{\kelvin} and above.
		}\label{app_fig:6}
	\end{minipage}\par

	\makeatletter
\section*{References}
	\begingroup
	\@fileswfalse
	\let\SMoriginalbibliography\NAT@thebibliography
	\let\SMoriginalendbibliography\endNAT@thebibliography
	\let\bibsection\@empty
	\long\def\SMfilteritem[#1]#2{\SMfilterbody{#2}}
	\long\def\SMfilterbody#1\BibitemOpen#2\BibitemShut#3{%
		\ifcsname SM@cited@#1\endcsname
			\ifcsname b@#1\@extra@b@citeb\endcsname
				\item[\@biblabel{\NAT@parse{#1}\NAT@num}]%
			\else
				\item[\@biblabel{?}]%
			\fi
			\Hy@raisedlink{\hyper@anchorstart{cite.SM.#1}\hyper@anchorend}%
			\BibitemOpen#2%
			\ifstrequal{#3}{NoStop}{.}{%
				\ifstrequal{#3}{Stop}{}{%
					\PackageError{SMrefs}{Unsupported bibliography ending: #3}{}%
				}%
			}%
		\fi
	}
	\expandafter\def\csname thebibliography\endcsname#1{%
		\SMoriginalbibliography{#1}%
		\let\bibitem\SMfilteritem
		\let\label\@gobble
	}
	\let\endthebibliography\SMoriginalendbibliography
	\IfFileExists{\jobname.bbl}{
}{}
	\endgroup
	\makeatother
\end{document}